\documentclass[conference,a4paper]{IEEEtran}

\usepackage[utf8]{inputenc} 
\usepackage[T1]{fontenc}
\usepackage{url}              
\usepackage{cite}             

\usepackage[cmex10]{amsmath}  
\usepackage{mleftright}       
\mleftright                   

\usepackage{graphicx}         
\usepackage{booktabs}         

\usepackage{tikz}
\usetikzlibrary{calc,decorations.pathmorphing,shapes}
\usepackage{xcolor}
\usepackage{colortbl} 
\usepackage{mathdots} 
\usepackage{amsfonts} 
\usepackage{mathabx,mathtools}
\usepackage{enumitem}
\usepackage{booktabs}
\usepackage{multirow}
\usepackage{arydshln}
\usepackage{adjustbox}
\usepackage{pgfplots}
\usepackage{blkarray}
\usepackage{tablefootnote}
\newtheorem{theorem}{Theorem}
\newtheorem{definition}{Definition}

\newtheorem{lemma}[theorem]{Lemma}
\newtheorem{corollary}[theorem]{Corollary}

\newtheorem{example}{Example}[section]

\newcommand{\x}{\Asterisk}
\newcommand{\nix}[1]{}
\newcommand{\B}[1]{#1}

\newlength{\rowidth}
\AtBeginDocument{\setlength{\rowidth}{4em}}

\definecolor{red2}{HTML}{FF003F}
\definecolor{teal2}{HTML}{009988}
\definecolor{cyan2}{HTML}{33BBEE}
\definecolor{yel2}{HTML}{EECC66}
\definecolor{purp2}{HTML}{AA4499}

\newcounter{rarrow}
\newcommand\rbc[1]{%
\stepcounter{rarrow}%
\mathrel{\begin{tikzpicture}[baseline= {( $ (current bounding box.south) + (0,-0.3ex) $ )}]
\node[font=\fontsize{5}{0}\selectfont,inner sep=.5ex] (\therarrow) {$\scriptstyle #1$};
\path[draw,<-,decorate,
  decoration={snake,amplitude=0.7pt,segment length=1.2mm,pre=lineto,pre length=4pt}] 
    (\therarrow.south east) -- (\therarrow.south west);
\end{tikzpicture}}%
}

\newcounter{larrow}
\newcommand\lbc[1]{%
\stepcounter{larrow}%
\mathrel{\begin{tikzpicture}[baseline= {( $ (current bounding box.south) - (0,-2ex) $ )}]
\node[font=\fontsize{5}{0}\selectfont,inner sep=.5ex] (\thelarrow) {$\scriptstyle #1$};
\path[draw,<-,decorate,
  decoration={snake,amplitude=0.7pt,segment length=1.2mm,pre=lineto,pre length=4pt}] 
    (\thelarrow.north west) -- (\thelarrow.north east);
\end{tikzpicture}}%
}

\begin{document}

\title{Generalizations and Extensions to \\Lifting Constructions for Coded Caching} 


%
 \author{%
   \IEEEauthorblockN{V. R. Aravind, Pradeep Kiran Sarvepalli, and Andrew Thangaraj, Senior Member, IEEE}
   \IEEEauthorblockA{Department of Electrical Engineering, Indian Institute of Technology Madras, India
                     }
                   }

\maketitle

\begin{abstract}
Coded caching is a technique for
achieving increased throughput in cached networks during peak
hours. Placement delivery arrays (PDAs) capture both placement and delivery scheme requirements in coded caching in a single array. Lifting is a method of constructing PDAs, where entries in a small base PDA are replaced with constituent PDAs that satisfy a property called Blackburn-compatibility.
We propose two new constructions for Blackburn-compatible PDAs
including a novel method for lifting Blackburn-compatible PDAs to obtain new sets of Blackburn-compatible PDAs.
Both of these constructions improve upon previous tradeoffs between rate, memory and subpacketization.
We generalize lifting constructions by defining partial Blackburn-compatibility between  two PDAs w.r.t. a third PDA.
This is a wider notion of Blackburn-compatibility making the original definition a special case. 
We show that some popular coded caching schemes can be defined as lifting constructions in terms of this extended notion.


\end{abstract}

\section{Introduction}
\label{sec:intro}
Caching is a useful technique to reduce network traffic by storing files partly or fully, at or near the users, 
during off-peak hours. So, when the users demand files during the peak hours, only the remaining part of these files need to be transmitted. This reduction in communication load (also referred to as \textit{rate}) by caching files or parts of files is referred to as \textit{local caching gain}. Maddah-Ali and Niesen have demonstrated in \cite{maddah2014fundamental} that the use of coding can achieve a gain (referred to as \textit{coding gain} or \textit{global caching gain}) additional to the caching gain in cached broadcast networks. They also proposed a centralized scheme which is order-optimal with respect to an information-theoretic lower bound on the rate. An improved version of this scheme from \cite{yu2017exact} was found to be optimal for uncoded prefetching.

\nix{
Coded caching is an active research area attracting extensive interest. 
It has been studied in many scenarios including decentralised caching \cite{maddah2015decentralized}, online caching \cite{pedarsani2015online}, coded prefetching \cite{tian2018caching}, hierarchical topology \cite{karamchandani2016hierarchical}, device-to-device networks \cite{ji2015fundamental}, non-uniform demands \cite{niesen2016coded}, content security \cite{sengupta2014fundamental,ravindrakumar2016fundamental} and demand privacy \cite{wan2020coded,kamath2019demand,aravind2020subpacketization,yan2020fundamental}. There have been several works focused on characterizing the rate-memory tradeoff in coded caching by improving the lower bounds \cite{yu2018characterizing,ghasemi2017improved,sengupta2015improved} and achievable schemes \cite{amiri2016fundamental,tian2018caching,yu2017exact}. 
}
Coded caching is an active research area that has been studied extensively \cite{maddah2015decentralized,pedarsani2015online,tian2018caching,karamchandani2016hierarchical,ji2015fundamental,niesen2016coded,sengupta2014fundamental,ravindrakumar2016fundamental,wan2020coded,aravind2020subpacketization,yan2021fundamental,yu2018characterizing,ghasemi2017improved,sengupta2015improved,amiri2016fundamental,yu2017exact}.
One drawback of the schemes from \cite{maddah2014fundamental,yu2017exact} is high subpacketization when the number of users is large. 
Subpacketization is the number of subfiles a file needs to be divided into for the purpose of coding. High subpacketization necessitates files to be large and also entails high computational overhead. There is an ongoing interest in characterizing the rate-memory tradeoff for coded caching schemes with low subpacketization. Towards this end, several combinatorial frameworks have been proposed \cite{yan2017placement,yan2017bipartite,shangguan2018centralized,shanmugam2017unified} to design coded caching schemes.
The framework of placement delivery arrays  (PDAs) was proposed by Yan \textit{et al.} in \cite{yan2017placement} for designing both placement and delivery schemes in coded caching using an array. Other combinatorial frameworks for designing coded caching schemes include bipartite graphs \cite{yan2017bipartite}, $3$-partite hypergraphs \cite{shangguan2018centralized}, Ruzsa-Szeméredi graphs \cite{shanmugam2017unified} \textit{etc.} Other coded caching schemes with lower subpacketization include \cite{tang2018coded,krishnan2018coded,cheng2020some,cheng2021framework,huang2021new,zhong2020placement,aravind2022lifting}.
However, the exact pareto-optimality between memory, rate, and subpacketization remains an open problem \cite{cheng2017coded}.

Designing PDAs for low subpacketization is 
hard. 
Combining existing PDAs to obtain new PDAs \cite{michel2019placement,zhong2020placement,aravind2022lifting} provides a good framework for constructing PDAs.
%
In \cite{michel2019placement}, the authors proposed constructions for PDAs by taking products of the underlying strong-edge colored bipartite graphs of existing PDAs. Concatenation constructions of PDAs were proposed in \cite{zhong2020placement} using a similar approach. In \cite{aravind2022lifting}, the authors proposed a framework for lifting constructions in terms of Blackburn-compatible PDAs. 
Several constructions for Blackburn-compatible PDAs were also proposed in \cite{aravind2022lifting}.
In lifting constructions, entries in a base PDA are replaced with Blackburn-compatible PDAs of the same size to obtain a new larger PDA. 

In this paper, we propose new constructions of PDAs and generalize the notion of Blackburn-compatibility for lifting constructions. Our new constructions improve upon the tradeoff between rate, memory, and subpacketization. With the generalization, we could bring some of the important coded caching schemes into the ambit of lifting constructions.
Our main contributions are summarized below.
\begin{enumerate}
    \item We propose a construction of square Blackburn PDAs whose size can be any odd number, including prime numbers. Using this, we could obtain PDAs for number of users with very few factors. Hence we extended the range of lifting constructions.
    \item We propose a novel method for obtaining sets of Blackburn-compatible PDAs from existing sets. Using this method in conjunction with some of the constructions from \cite{aravind2022lifting}, we show the achievability of improved memory-rate tradeoffs.
    \item We generalize the idea of lifting to include constructions which previously could not be explained in terms of lifting. In particular, we show that the schemes from \cite{maddah2014fundamental,shangguan2018centralized,yan2017placement} can be obtained using lifting constructions. 
\end{enumerate}

The paper is organized as follows. Section~\ref{sec:prelims} introduces the problem setup, background, and notations. 
In Section~\ref{sec:construction}, we propose new constructions for Blackburn-compatible PDAs. 
In Section~\ref{sec:Blackburn}, we define the new notion of left and right Blackburn-compatibility between two PDAs with respect to other PDAs. We also define the non-uniform lifting construction for PDAs. 
We discuss numerical results from our constructions and compare them with existing schemes in Section~\ref{sec:results}.

\section{Preliminaries}\label{sec:prelims}
\subsection{Problem Setup}
\begin{figure}[htb]
     \centering
     \includegraphics[scale=1.25]{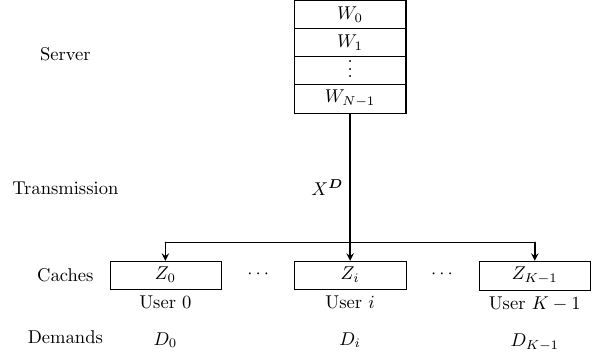}
    \caption{A broadcast network with users with distinct caches. A server with $N$ files connected to $K$ users, each with a cache memory of $MF$ bits.}
     \label{fig:setup}
 \end{figure}

Assume we have a server connected to $K$ users over a broadcast link. The server has $N$ files, denoted by $W_i, i\in [N]$, of size $F$ bits. Each user has a cache memory of $MF$ bits denoted by $Z_k, k\in [K]$. The system setup is shown in Fig.~\ref{fig:setup}.
\vspace{-1.8em}

During the off-peak hours or the \textit{placement phase}, the caches are populated with parts or functions of the files. During the peak-hours or \textit{delivery phase}, each user demands a single file from the server. Demand of User~$k$ is denoted by $D_i\in[N]$ and $\boldsymbol{D}=(D_k: k\in[K])$ is called the demand vector. The server sends a transmission of $RF$ bits denoted by $X^{\boldsymbol{D}}$. For each $k\in[K]$, User~$k$ should be able to recover the file it demanded ($W_{D_k}$) from this transmission ($X^{\boldsymbol{D}}$) and its cache contents ($Z_k$).

\subsection{Placement Delivery Arrays}
When the cache prefetching is uncoded and symmetric with respect to all files, the placement and delivery schemes can be represented in a single array \cite{yan2017placement}.
\begin{definition}[Yan \textit{et al.} \cite{yan2017placement}]
    For positive integers $K$, $f$, $Z$ and a set of integers $\mathcal{S}$, a $(K, f, Z, \mathcal{S})$ \textit{placement delivery array} is an $f\times K $ matrix  $P=[p_{j,k}]$, $j\in[f],k\in[K]$, containing either a ``$\x$'' or integers from $\mathcal{S}$ in each cell such that they satisfy the following conditions.
\begin{enumerate}[label=C\arabic*., ref=C\arabic*]
    \item (Memory constraint) The symbol $\x$ appears $Z$ times in each column.\label{cond:equalZ}
    \item (Rate constraint) Each integer $s\in \mathcal{S}$ occurs at least once in the array.\label{cond:everysOnce}
    \item (Blackburn property) If the entries in two distinct cells $p_{j_1,k_1}$ and $p_{j_2,k_2}$ are the same integer $s\in\mathcal{S}$, then $p_{j_{1},k_{2}}=p_{j_{2},k_{1}}=\Asterisk$.\label{cond:blackburn}
\end{enumerate}
\end{definition}

A $(K,f,Z,\mathcal{S})$ PDA, $P = [p_{j,k}]_{f\times K}$,
represents a coded caching system with $K$ users, subpacketization $f$, $M/N = Z/f$ and $R = |\mathcal{S}|/f$ \cite{yan2017placement}.
 A $(K,f,Z,\mathcal{S})$ PDA $P$ is said to be $g$-regular and denoted $g$-$(K,f,Z,\mathcal{S})$ if the coding gain is $g$, \textit{i.e.}, if each integer in $\mathcal{S}$ appears $g$ times in $P$.

For a set $\mathcal{N}$ and integer $n$, the collection of subsets $\{\mathcal{M}:\mathcal{M}\subseteq\mathcal{N}, |\mathcal{M}|=n\}$ is denoted by $\binom{\mathcal{N}}{n}$. The set of integers $\{0,1,\ldots,n-1\}$ is denoted by $[n]$. We denote a $(K,f,Z,[m])$ PDA with  $\mathcal{S}=[m]$ as a $(K,f,Z,m)$ PDA.

\subsection{Commonly Used PDAs}
Notations for some basic PDAs are as follows. 
\begin{enumerate}
    \item For an integer $s$, $I_n(s)$ denotes the $(n,n,n-1,1)$ PDA with the integer $s$ on the main diagonal and $\x$ in all other cells. $\tilde{I}_n(s)$ denotes the $(n,n,n-1,1)$ PDA with the integer $s$ on the main anti-diagonal and $\x$ in all other cells \cite{maddah2014fundamental}.
    For example, 
    \begin{align}
    \resizebox{0.65\columnwidth}{!}{$
    I_3(1)=\begin{pmatrix}
    1  & \x & \x \\
    \x & 1  & \x \\
    \x & \x & 1
    \end{pmatrix},\quad \tilde{I}_3(0)=\begin{pmatrix}
    \x & \x & 0  \\
    \x & 0  & \x \\
    0  & \x & \x
    \end{pmatrix}.$}\label{eq:It}
    \end{align}
    \item For a sequence $\mathcal{S}=\{s_1,\ldots,s_{m}\}$, $m=n(n-1)/2$, the following are $2$-$(n,n,1,n(n-1)/2)$ PDAs \cite{maddah2014fundamental}:
\begin{align}
    \resizebox{0.83\columnwidth}{!}{$
    G_n(\mathcal{S})=\begin{pmatrix}
    s_1  & \cdots  &s_{n-1} &\x \\
    \vdots&\iddots&\iddots&s_{n-1}\\
    s_{m}  & \x & \iddots&\vdots  \\
    \x & s_{m}  & \cdots&s_1
    \end{pmatrix},\, 
H_n(\mathcal{S})=\begin{pmatrix}
    \x & s_1 & \cdots & s_{n-1} \\
    s_1 & \ddots & \ddots &\vdots \\
    \vdots & \ddots & \x & s_m\\
    s_{n-1}&\cdots&s_m&\x
    \end{pmatrix}.$}\label{eq:HS}
    \end{align}
    \item For a sequence $\mathcal{S}=\{s_1,\ldots,s_{mn}\}$ of $mn$ integers, $J_{m,n}(\mathcal{S})$ denotes the $(n,m,0,\mathcal{S})$ PDA obtained by filling all the cells in the array with distinct integers from $\mathcal{S}$ row-wise in the specified order. For example,
    \begin{align}
    J_{2,3}([9])=\begin{pmatrix}
    0  & 1  & 2 \\
    3  & 4  & 5 \\
    \end{pmatrix}.\label{eq:JS}
    \end{align}
    \item $\x_{m,n}$ denotes the $(n,m,m,\emptyset)$ PDA obtained by filling all the cells in the array with $\x$. For example,
    \begin{align}
    \x_{2,3}=\begin{pmatrix}
    \x  & \x  & \x \\
    \x  & \x  & \x
    \end{pmatrix}.\label{eq:JS}
    \end{align}
    \item $M_{K,t}(\mathcal{S})$ (or MN PDA) denotes the  PDA of the Maddah-Ali-Niesen scheme \cite{maddah2014fundamental} for $K$ users and $\frac{M}{N}=\frac{t}{K}$ \cite{maddah2014fundamental,yan2017placement}. Let $M_{K,t}(\mathcal{S})$ be a $\binom{K}{t}\times K$ array. Columns of $M_{K,t}(\mathcal{S})$ are denoted by elements in $[K]$ and rows are denoted by sets in $\{\mathcal{T} \subset [K]:|\mathcal{T}|=t\}$ in lexicographic order. 
    Let $\sigma$ be a mapping from sets in  $\{\mathcal{U} \subset [K]:|\mathcal{U}|=t+1\}$ to their indices when listed in lexicographic order. 
    Then the entry $m_{\mathcal{T},k}$ in row $\mathcal{T}$ and column $k$ is
    \begin{equation}
    m_{\mathcal{T}, k}= \begin{cases}*, & \text { if } k \in \mathcal{T}, \\ \sigma(\mathcal{T} \cup\{k\}), & \text { if } k \notin \mathcal{T}.\end{cases}
    \end{equation}
    Let $\Tilde{M}_{K,t}(\mathcal{S})$ be defined similarly except for rows and integers labelled in reverse lexicographic order. The following are examples:

\hspace{-6.5em}
    \resizebox{0.69\textwidth}{!}{
    \begin{minipage}{\textwidth}
\begin{align*}
\begin{blockarray}{ccccc}
&  0 & 1 & 2 & 3 \\
\begin{block}{c(cccc)}
    01 &\x & \x & 0 & 1 \\
    02 & \x & 0 & \x & 2 \\
    03 & \x & 1 & 2 & \x \\
    12 & 0 & \x & \x & 3 \\
    13 & 1 & \x & 3 & \x \\
    23 & 2 & 3 & \x & \x \\
\end{block}
\\
& \BAmulticolumn{4}{c}{M_{4,2}([4])}\\
\end{blockarray}
\qquad
\begin{blockarray}{ccccc}
&  0 & 1 & 2 & 3 \\
\begin{block}{c(cccc)}
    23 & 1 & 0 & \x & \x \\
    13 & 2 & \x & 0 & \x \\
    12 & 3 & \x & \x & 0 \\
    03 & \x & 2 & 1 & \x \\
    02 & \x & 3 & \x & 1 \\
    01 & \x & \x & 3 & 2 \\
\end{block}
\\
& \BAmulticolumn{4}{c}{\tilde{M}_{4,2}([4])}\\
\end{blockarray}
\end{align*}
\end{minipage}
}
\end{enumerate}

\subsection{Lifting Constructions}
A construction of PDAs called lifting, proposed in \cite{aravind2022lifting},
is based on the concept of Blackburn-compatibility.
\begin{definition}
    Two $n\times n$ PDAs $P_0=[p^{(0)}_{ij}]$ and $P_1=[p^{(1)}_{ij}]$ are said to be \emph{Blackburn-compatible} w.r.t. a third $n\times n$ PDA $P_{\x}=[p^{({\x})}_{ij}]$ if, whenever  $p^{(0)}_{i_0j_0}{=}p^{(1)}_{i_1j_1}{\ne}\Asterisk$ is a common integer, then the \textit{mirrored} locations $p^{({\x})}_{i_0j_1}{=}p^{({\x})}_{i_1j_0}=\Asterisk$.
\end{definition}
%

For $g\ge2$, PDAs $\{P_0,\ldots,P_{g-1}\}$ are said to be Blackburn-compatible w.r.t. $P_*$ when they are pairwise Blackburn-compatible w.r.t. $P_*$. 

Given a PDA $P$ with integer set $\mathcal{S}$, an integer-disjoint copy of $P$ is obtained by replacing the integers with a disjoint set $\tilde{\mathcal{S}}$ (i.e. $\mathcal{S}\cap \tilde{\mathcal{S}}=\emptyset$ and $|\mathcal{S}| = |\tilde{\mathcal{S}}|$). The idea of uniform lifting from the General Lifting Theorem of \cite{aravind2022lifting} uses integer-disjoint copying along with Blackburn-compatibility, and is reproduced below for convenience. 
\begin{theorem}[$m\times n$ uniform lifting \cite{aravind2022lifting}]\label{th:generalLifting}
Let $P_b$ be a $(K,f,Z_b,\mathcal{S}_b)$ base PDA with $s\in\mathcal{S}_b$ occurring $g_s$ times.
Let $\mathcal{P}=\{P_1,\ldots,P_g\}$, $g=\max_s g_s$, be a set of $(n,m,Z_c,\mathcal{S}_t)$ PDAs Blackburn-compatible w.r.t. an $(n,m,Z_*,\mathcal{S}_*)$ PDA $P_*$. 
Let $\{P_{s,t}:t\in[g_s]\}$ be integer-disjoint copies of $\{P_t:t\in[g_s]\}$ for each $s\in\mathcal{S}_b$. 
Let $P_{*,r}$, $r\in[KZ_b]$, be integer-disjoint copies of $P_*$, which are integer-disjoint with $P_{s,t}$ for all $s,t$.

The base PDA $P_b$ is lifted to $P\triangleq \mathcal{L}_{\mathcal{P},P_*}(P_b)$ as follows:
\begin{enumerate}
    \item $r$-th $\x$ in $P_b$ is replaced by $P_{\x,r}$ for $r\in[KZ_b]$.
    \item $t$-th occurrence of integer $s\in\mathcal{S}_b$ in $P_b$ is replaced by $P_{s,t}$ for $t=1,\ldots,g_s$.
\end{enumerate}
Then, $P$ is a $(Kn,fm,Z_bZ_*+(f-Z_b)Z_c,\mathcal{S})$ PDA, where $\mathcal{S}=\left(\bigcup_{r\in[KZ_b]}\mathcal{S}_{*,r}\right)\bigcup\left(\bigcup_{s\in\mathcal{S}_b}\bigcup_{t\in[g_s]}\mathcal{S}_{s,t}\right)$, and $S_{*,r}$ is the set of integers in $P_{*,r}$.
\end{theorem}
When $P_*=\x_{m,n}$, $\{P,P,\ldots\}$ for any PDA $P$ is Blackburn-compatible w.r.t. $\x_{m,n}$. $\mathcal{L}_{\{P,P,\ldots\},\x_{m,n}}(P_b)$ is termed as \textit{basic lifting} and denoted 
$\mathcal{L}_{P,\x}(P_b)$ in short \cite{aravind2022lifting}.

\section{New Constructions}\label{sec:construction} 
In this section, we present two new constructions for Blackburn-compatible PDAs and use them for constructing lifted PDAs. First, we propose a construction of two Blackburn-compatible PDAs with size as any odd number $g$ including $g$ being prime. Although Constructions C1 and T1 in \cite{aravind2022lifting} provide prime-number-sized PDAs, the lifting using these PDAs did not result in an increased coding gain. The proposed construction can lift a $2$-PDA to obtain a $g$-PDA. 
\begin{theorem}[Odd tiling]\label{th:odd}
    Let $g\ge3$ be an odd integer and $n=\lfloor \frac{g}{2}\rfloor$. Then the $(g,g,g-2,[4])$ PDAs $P_0$ and $P_1$ defined in \eqref{eq:odd} are Blackburn-compatible w.r.t. $I_g(s)$.
\begin{align}
\scalebox{0.8}{$
\begin{aligned}
    P_0=\left(\begin{array}{ccccccc}
\multicolumn{3}{c}{\multirow{2}{*}{$I_n(0)$}}      & \multicolumn{3}{c}{\multirow{2}{*}{$I_n(1)$}}      & \multirow{2}{*}{$\x_{n,1}$} \\
\multicolumn{3}{c}{}                           & \multicolumn{3}{c}{}                           &                    \\ \hdashline
2                            & \multicolumn{5}{c}{\x_{1,g-2}}                                            & 1                  \\ \hdashline
\multirow{2}{*}{$\x_{n,1}$}           & \multicolumn{3}{c}{\multirow{2}{*}{$I_n(2)$}} & \multicolumn{3}{c}{\multirow{2}{*}{$I_n(3)$}} \\
                             & \multicolumn{3}{c}{}                      & \multicolumn{3}{c}{}
    \end{array}\right),\,
    P_1=\left(\begin{array}{ccccccc}
\multirow{2}{*}{$\x_{n,1}$} & \multicolumn{3}{c}{\multirow{2}{*}{$\Tilde{I}_n(3)$}}      & \multicolumn{3}{c}{\multirow{2}{*}{$\Tilde{I}_n(1)$}}      \\
\multicolumn{3}{c}{}                           & \multicolumn{3}{c}{}                           &                    \\ \hdashline
3                            & \multicolumn{5}{c}{\x_{1,g-2}}                                            & 0                  \\ \hdashline
           \multicolumn{3}{c}{\multirow{2}{*}{$\Tilde{I}_n(2)$}} & \multicolumn{3}{c}{\multirow{2}{*}{$\Tilde{I}_n(0)$}} & \multirow{2}{*}{$\x_{n,1}$}\\
                             & \multicolumn{3}{c}{}                      & \multicolumn{3}{c}{}
    \end{array}\right).\label{eq:odd}
\end{aligned}$}
\end{align}
\end{theorem}

\begin{IEEEproof}
The set of integers in $P_0$ and $P_1$ is $[4]$. Consider the integer $0$, which occurs in both the PDAs $P_0$ and $P_1$. 
Then $p^{(0)}_{ij}=0$ implies $i,j\in[n]$ and  $p^{(1)}_{ij}=0$ implies $i,j\in[g]\setminus[n]$. 
Therefore, $p^{(0)}_{i_0j_0}{=}p^{(1)}_{i_1j_1}=0$ implies $p^{({\x})}_{i_0j_1}=\x$ since $i_0\in[n]$, $j_1\in[g]\setminus [n]$ and $p^{({\x})}_{ij}=\x, i,j\in[g]$ when $i\ne j$. 
Also, 
$p^{({\x})}_{i_1j_0}=\x$ since $j_0\in[n]$, $i_1\in[g]\setminus [n]$. It can be similarly shown for other integers also.
Hence, $P_0$ and $P_1$ are Blackburn-compatible w.r.t. $P_*$.
%
\end{IEEEproof}
This leads to the following lifting construction for a particular choice of the base PDA.
\begin{corollary}\label{co:odd}
    For integers $g,n$ such that $g$ is odd, there exists a $g$-regular $(gn,gn,n(g-2)+1,n(2n-1))$ PDA.
\end{corollary}
\begin{IEEEproof}
    Lifting $G_n(\mathcal{S})$ using the Blackburn-compatible PDAs from Theorem~\ref{th:odd} results in a PDA as claimed.
\end{IEEEproof}
\begin{example}
    For $n=2$ and $g=5$ we obtain the following $5$-regular $(10,10,7,6)$ PDA using Corollary~\ref{co:odd}.
    \begin{equation*}
    \resizebox{0.7\columnwidth}{!}{$
\left(\begin{array}{ccccc:ccccc}
0 & \x & \x & \x & \x & 2 & \x & 3 & \x & \x \\
\x & 0 & \x & \x & \x & \x & 2 & \x & 3 & \x \\
\x & \x & 0 & \x & \x & 4 & \x & \x & \x & 3 \\
\x & \x & \x & 0 & \x & \x & 4 & \x & 5 & \x \\
\x & \x & \x & \x & 0 & \x & \x & 4 & \x & 5 \\ \hdashline
\x & \x & 5 & \x & 3 & 1 & \x & \x & \x & \x \\
\x & 5 & \x & 3 & \x & \x & 1 & \x & \x & \x \\
5 & \x & \x & \x & 2 & \x & \x & 1 & \x & \x \\
\x & 4 & \x & 2 & \x & \x & \x & \x & 1 & \x \\
4 & \x & 2 & \x & \x & \x & \x & \x & \x & 1
\end{array}\right)
    $}
    \end{equation*}
\end{example}

In the recursive construction of \cite[Lemma~11]{aravind2022lifting}, a set of Blackburn-compatible PDAs are lifted to obtain a set of larger Blackburn-compatible PDAs. In our second construction, we generalize this approach in the following lemma.
\begin{lemma}[Lifting Blackburn-compatible PDAs]\label{lm:liftbc}
Let $\mathcal{P}=\{P_0,P_1,\ldots,P_{d-1}\}$ be a set of $g$-PDAs with a common integer set, Blackburn-compatible w.r.t. a $g$-PDA $P_*$. 
Let $\mathcal{Q}=\{Q_0,Q_1,\ldots,Q_{g-1}\}$ be Blackburn-compatible w.r.t. $Q_*$. 

Let $R_*=\mathcal{L}_{Q_0,\x}(P_*)$ be a PDA created by basic lifting of $P_*$ using $Q_0$. Let $R_i=\mathcal{L}_{\mathcal{Q},Q_*}(P_i)$, $i\in[d]$. Then $\mathcal{R}=\{R_i:i\in[d]\}$ is a set of Blackburn-compatible PDAs w.r.t. $R_*$, if the following condition is satisfied: 

[C$\x$] If $Q_*(\mathcal{S})$ with the same integer set $\mathcal{S}$ replaces $\x$s at $(r_i,c_i)$ in $P_i$ and $(r_j,c_j)$ in $P_j$ for $i\ne j$, then  $P_*(r_i,c_j)=P_*(r_j,c_i)=\x$. 
\end{lemma}
\begin{IEEEproof}
Each PDAs in $\mathcal{P}$ has the same set of integers. Therefore, PDAs that replaced $s_i$ in $P_i$ and $s_j$ in $P_j$ has same integers only if $s_i=s_j$. Since the mirrored locations in $P_*$ is replaced by an all-$\x$ array through basic lifting, Blackburn property holds for these integers. Blackburn property holds for the integers in PDAs that replaces $\x$s in $P_i$ because of the Condition C$\x$ given in the Lemma. Hence, $\mathcal{R}$ is a set of Blackburn-compatible PDAs w.r.t. $R_*$.
\end{IEEEproof}

For the Blackburn-compatible PDAs obtained from constructions C2, BW1, BW2, BW3 in \B{\cite[Lemmas~4, 6, 7 and 8]{aravind2022lifting}}, the cells in $P_i$s with $\x$s occur in the diagonal and the mirrored locations in the diagonal of $P_*$ also have $\x$s in them. Hence, Condition C$\x$ is satisfied and lifting using Lemma~\ref{lm:liftbc} is possible.

We show below an example for the constructions using Lemma~\ref{lm:liftbc}. 
We use the notation $(K,f)_Z^g$ from \cite{aravind2022lifting} for a $g$-$(K,f,Z,\mathcal{S})$ PDA. Let $\mathcal{P}=\{P_i: i\in[g_b]\}$ be a set of $(K,f,Z_i,\mathcal{S}_i)$ PDAs Blackburn-compatible w.r.t. a $g_L$-regular $(K,f,Z_*,\mathcal{S}_*)$ PDA $P_*$. We denote $\{\mathcal{P},P_*\}$ with $(K,f)_{Z_i,Z_*}^{g_b,g_L}$. 
\begin{example}
Let $B(g,d)$ denote the set of Blackburn-compatible PDAs obtained from Construction BW3 \cite{aravind2022lifting}. A new set of Blackburn-compatible PDAs can be obtained by lifting $B(6,3)$ with $B(10,2)$ using Lemma~\ref{lm:liftbc} as follows.
\begin{equation}
    B(6,3)=(6,6)_{1,5}^{3,6}\xrightarrow[Lemma~\ref{lm:liftbc}]{B(10,2)=(10,10)_{1,6}^{2,4}} (60,60)_{11,51}^{3,12}\label{eq:bw3pda}
\end{equation}
{Using the above Blackburn-compatible PDAs}, we can lift any 3-PDA to obtain a 12-PDA. An example is shown below.
\begin{equation*}
    M_{4,2}([4])=(4,6)_{3}^{3}\xrightarrow[Lifting~(Th.~\ref{th:generalLifting})]{(60,60)_{11,51}^{3,12}} (240,360)_{186}^{12}.
\end{equation*}
A few more examples are listed below.
The following constructions for PDAs use Blackburn-compatible PDAs from Constructions BW3 and $2^r$-lifting in \cite{aravind2022lifting} in conjunction with Lemma~\ref{lm:liftbc}.
\begin{align*}
    &(5,10)_{4}^{3}\xrightarrow[Lifting~(Th.~\ref{th:generalLifting})]{(6,6)_{1,5}^{3,6}\xrightarrow[Lemma~\ref{lm:liftbc}]{(8,8)_{1,5}^{2,4}} (48,48)_{10,41}^{3,12}} (240,480)_{224}^{12} \\
    &(5,5)_{1}^{2}\xrightarrow[Lifting~(Th.~\ref{th:generalLifting})]{(6,6)_{1,4}^{2,4}\xrightarrow[Lemma~\ref{lm:liftbc}]{(8,8)_{1,5}^{2,4}} (48,48)_{10,34}^{2,8}} (240,240)_{74}^{8} \\
    &(5,10)_{4}^{3}\xrightarrow[Lifting~(Th.~\ref{th:generalLifting})]{(3,3)_{1,3}^{3,6}\xrightarrow[Lemma~\ref{lm:liftbc}]{(16,16)_{6,13}^{2,8}} (48,48)_{25,48}^{3,24}} (240,480)_{342}^{24} \\
    &(4,4)_{1}^{2}\xrightarrow[Lifting~(Th.~\ref{th:generalLifting})]{(4,4)_{1,3}^{2,4}\xrightarrow[Lemma~\ref{lm:liftbc}]{(16,16)_{6,13}^{2,8}} (64,64)_{31,54}^{2,16}} (256,256)_{147}^{16}
\end{align*}
\end{example}

\section{Generalizing Blackburn-Compatibility}\label{sec:Blackburn}
One aspect of the uniform lifting constructions is that the size of the lifted PDA is proportional to the size of the Blackburn-compatible PDAs.
This motivates alternative nonuniform constructions where the size could change differently.
For this, the definition of Blackburn-compatibility needs to be generalised, so that entries in the base PDA may be lifted by PDAs of different sizes to break the symmetry in uniform lifting.
\subsection{Partial Blackburn-Compatibility}
We introduce partial Blackburn-compatibility for PDAs.
Using this definition, we develop a new strategy for lifting PDAs.

\nix{We present a generalized notion of Blackburn-compatibility in this section. In \cite{aravind2022lifting}, two PDAs were defined to be Blackburn-compatible when there was  We define Blackburn-compatibility between rows of a PDA to columns of another PDA with respect to a third PDA. 
There are two major implications to this new definition.
\begin{enumerate}
    \item Blackburn-compatibility can be defined between PDAs of different sizes.
    \item Rows and columns of a PDA can be Blackburn-compatible to the columns and rows respectively of another PDA w.r.t. two completely different PDAs.
\end{enumerate}}

\begin{definition}
Let $P_0=[p^{(0)}_{ij}]$ be an $m_0\times n_0$ PDA and $P_1=[p^{(1)}_{ij}]$ be an $m_1\times n_1$ PDA.  The PDAs $P_0$ and $P_1$ are said to be \textit{right Blackburn-compatible} (in that order) w.r.t. an $m_0\times n_1$ PDA $P_{\x}=[p^{({\x})}_{ij}]$ if $p^{(0)}_{i_0j_0}=p^{(1)}_{i_1j_1}\ne \x$ implies $p^{({\x})}_{i_0j_1}=\x$. Similarly, $P_0$ and $P_1$ are \textit{left Blackburn-compatible} w.r.t. an $m_1\times n_0$ PDA $P_\#=[p^{({\#})}_{ij}]$ if $p^{(0)}_{i_0j_0}=p^{(1)}_{i_1j_1}\ne \x$ implies $p^{({\#})}_{i_1j_0}=\x$.
\end{definition}

We use the notation $P_0 \rbc{P_*^{0,1}} P_1$ to denote right Blackburn-compatibility between $P_0$ and $P_1$ w.r.t. $P_*^{0,1}$.
Left Blackburn-compatibility between $P_0$ and $P_1$ w.r.t. $P_*^{1,0}$ is denoted by $P_0 \lbc{P_*^{1,0}} P_1$.
Note that $P_0 \lbc{P_*^{1,0}} P_1$ is equivalent to $P_1 \rbc{P_*^{1,0}} P_0$.
Let $P_0$, $P_1$ and $P_*$ be of the same size. 
If $P_0$ and $P_1$ are both left and right Blackburn-compatible w.r.t. the same PDA $P_*$, then we fall back to the old definition of PDAs $P_0$ and $P_1$ being {Blackburn-compatible} w.r.t. PDA $P_*$ as in \cite{aravind2022lifting}.

A set of PDAs $\mathcal{P}=\{P_i :  i\in [g]\}$ are said to be \textit{generalized Blackburn-compatible} w.r.t. another set of PDAs $\mathcal{P}_*=\{P_*^{(i,j)}: i,j\in[g], i\ne j\}$ when $P_i$ and $P_j$ are right Blackburn-compatible w.r.t. $P_*^{(i,j)}$ and left Blackburn-compatible w.r.t. $P_*^{(j,i)}$.
{In the special case when $P_*^{i,j}=P_* \text{ for all } i,j\in[g]$, generalized Blackburn-compatibility reduces to the earlier definition \cite{aravind2022lifting}.}
\begin{theorem}[Non-uniform lifting of an identity PDA]\label{th:nonuni}
Suppose $\mathcal{P}=\{P_i : i\in [g]\}$ are $(n_i,m_i,Z_i,\mathcal{S}_i)$ PDAs. Let $\mathcal{P}_*=\{P_*^{(i,j)}: i,j\in[g], i\ne j\}$ be $(n_{i,j},m_{i,j},Z_*^{(i,j)},\mathcal{S}_*^{(i,j)})$ PDAs containing integers that are disjoint from each other and from the integers in $P_0,\ldots,P_{g-1}$. Let $Z=Z_j+\sum_{i\in[g]\setminus j} Z*^{(i,j)}$ be a constant for all $j\in[g]$. Then, 
for $i\ne j$, $P_i$ and $P_j$ are right Blackburn-compatible w.r.t. $P_*^{(i,j)}$
if and only if the following lifting of $I_g$ is a valid $(\sum_i n_i,\sum_i m_i,Z,\mathcal{S})$ PDA: 
\begin{equation}
\nix{\mathcal{L}_{\mathcal{P},\mathcal{P}_*}(I_g)\triangleq\begin{pmatrix}
    P_0&\cdots&P_*^{(0,i)}&\cdots&P_*^{(0,g-1)}\\
    \vdots&&\vdots&&\vdots\\
    P_*^{(i,0)}&\cdots&P_i&\cdots&P_*^{(i,g-1)}\\
    \vdots&&\vdots&&\vdots\\
    P_*^{(g-1,0)}&\cdots&P_*^{(g-1,i)}&\cdots&P_{g-1}
\end{pmatrix}.}
{\mathcal{L}_{\mathcal{P},\mathcal{P}_*}(I_g)\triangleq\begin{pmatrix}
    P_0&P_*^{(0,1)}&\cdots&P_*^{(0,g-1)}\\
    P_*^{(1,0)}&P_1&\cdots&P_*^{(1,g-1)}\\
    \vdots&\vdots&\ddots&\vdots\\
    P_*^{(g-1,0)}&P_*^{(g-1,1)}&\cdots&P_{g-1}
\end{pmatrix}.
}\end{equation}
\end{theorem}
\begin{IEEEproof}
Let $Q=L_{\mathcal{P},\mathcal{P}_*}(I_g)$. By hypothesis $Z=Z_j+\sum_{i\in[g]\setminus j} Z*^{(i,j)}$, \ref{cond:equalZ} is satisfied. Since, the PDAs in $\mathcal{P}$ and $\mathcal{P}_*$ satisfy \ref{cond:everysOnce} and $\mathcal{S}= (\bigcup_{i\in[g]} \mathcal{S}_i) \bigcup (\bigcup_{i,j\in[g],i\ne j} \mathcal{S}_*^{i,j})$, \ref{cond:everysOnce} is satisfied for $Q$. PDAs $\mathcal{P}$ satisfying Blackburn-compatibility is equivalent to $Q$ satisfying \ref{cond:blackburn}. Hence the claim is proved.
\end{IEEEproof}

\nix{\begin{example}
    $\mathcal{P}=\{P_0,P_1,P_2\}$ given below is a set of Blackburn-compatible PDAs w.r.t. $\mathcal{P}_*=\{P_*^{(i,j)}:i,j\in[3],i\ne j\}$.
    \begin{align*}
        P_0=J_{2,1}([2]),\qquad P_1=&H_3([3]),\qquad P_2=P_0^T,\\
        P_*^{(0,1)}=\x_{2,3},\qquad P_*^{(0,2)}=&H_2{(1)},\qquad P_*^{(1,0)}=\x_{3,1},\\
        P_*^{(1,2)}=\x_{3,2},\qquad P_*^{(2,0)}=&\x_{1,1},\qquad P_*^{(2,1)}=\x_{1,3}.
    \end{align*}
    Lifting $I_3$ with using this set of Blackburn PDAs result in the following $4$-PDA,
    \begin{align*}
        \mathcal{L}_{\mathcal{P},\mathcal{P}_*}=
        \left(\begin{array}{c:ccc:cc}
        0 & \x & \x & \x & \x & 2 \\
        1 & \x & \x & \x & 2 & \x \\ \hdashline
        \x & \x & 0 & 1 & \x & \x \\
        \x & 0 & \x & 2 & \x & \x \\
        \x & 1 & 2 & \x & \x & \x \\ \hdashline
        \x & \x & \x & \x & 0 & 1
        \end{array}\right).
    \end{align*}
\end{example}}
\begin{example}
    $\mathcal{P}=\{P_0=\begin{pmatrix}
            I_2(0)\\
            I_2(1)
        \end{pmatrix},P_1=\begin{pmatrix}
            I_2(1) &
            I_2(0)
        \end{pmatrix}\}$ 
        is a set of generalized Blackburn-compatible PDAs w.r.t. $\mathcal{P}_*=\{P_*^{(0,1)}=I_4(1), P_*^{(1,0)}=\x_{2,2}\}$.
\nix{    \begin{align*}
        P_0=\begin{pmatrix}
            I_2(0)\\
            I_2(1)
        \end{pmatrix},\qquad 
        P_1=\begin{pmatrix}
            I_2(1) &
            I_2(0)
        \end{pmatrix}.
    \end{align*}}%
    Lifting $I_2$ using the above set of PDAs results in the following $4$-PDA:
    \begin{align*}
    \resizebox{0.6\columnwidth}{!}{$
        \mathcal{L}_{\mathcal{P},\mathcal{P}_*}=
        \left(\begin{array}{cc:cccc}
        0 & \x & 2 & \x & \x & \x \\
        \x & 0 & \x & 2 & \x & \x \\ 
        1 & \x & \x & \x & 2 & \x \\
        \x & 1 & \x & \x & \x & 2 \\ \hdashline
        \x & \x & 1 & \x & 0 & \x \\
        \x & \x & \x & 1 & \x & 0
        \end{array}\right).
        $}
    \end{align*}
\end{example}
The notion of partial Blackburn-compatibility allows more flexibility since compatibility between PDAs can be made w.r.t. different PDAs. Thus, we could have more control over the PDA parameters obtained from the lifting construction using Theorem~\ref{th:nonuni}. However, note that Theorem~\ref{th:nonuni} has a limitation on the choice of base PDAs we can have. Next, we show that some of the popular coded caching schemes can be cast as lifting constructions with this new framework.

\nix{A Blackburn-compatible set of PDAs $P_i, i\in[g]$ can be used to lift $I_g$ even when they are of different size.
But, in a typical base PDA, a $\x$ entry could be a mirrored location for many integers. In such a scenario, lifting using a Blackburn-compatible set of PDAs of different sizes might not be feasible.
Consider the case when two occurrences of two integers have mirrored location at the same cell in the base PDA. Let the two occurrences of one of the integers be replaced with two PDAs from a set of Blackburn-compatible PDAs. A unique choice of $P_*^{i_1,i_2}$ to replace the $\x$ at the mirrored location might not exist if two other PDAs from the set replace the two occurrences of the other integer.
}

\subsection{\B{Towards a} unified framework for existing constructions}


Some important existing constructions, including the MN PDAs, can be described as lifting constructions. 
The following lemma will lead to that description.
\begin{lemma}\label{lm:mnrec}
    The PDAs $J_{\binom{K}{t+1},1}(\mathcal{S})$ and $M_{K,t}(\mathcal{S})$ are right Blackburn-compatible w.r.t. $M_{K,t+1}(\mathcal{S_*})$ and left Blackburn-compatible w.r.t. $\x_{\binom{K}{t},1}$.
\end{lemma}
\begin{IEEEproof}
    Let $P_0=J_{\binom{K}{t+1},1}(\mathcal{S})$, $P_1=M_{K,t}(\mathcal{S})$ $P_*=M_{K,t+1}(\mathcal{S_*})$ and $P_\#=\x_{\binom{K}{t},1})$. 
    PDAs $P_0$ and $P_1$ are left Blackburn-compatible w.r.t. $P_\#$ since $P_\#$ is an all-$\x$ array.
    Rows in both $P_1$ and $P_*$ are labelled with $\binom{[K]}{t+1}$. Integers in $P_0$ and $P_1$ are also labelled with $\binom{[K]}{t+1}$ such that entries in $P_0$ and their rows have the same labels. Let $s$ be an integer with label $\sigma$ in $P_0$ and $P_1$. Let $\Gamma$ be the set of labels of columns containing integer $s$ in $P_1$. We have $\gamma\subset\sigma$ for all $\gamma\in\Gamma$ since the label of an integer is the union of its row and column labels. Since $\gamma\subset\sigma$, we also have $P_*(\sigma,\gamma)=\x$ for all $\gamma\in\Gamma$. 
\end{IEEEproof}
Using this result, we can define the MN construction as a recursive lifting construction, as shown below.

\begin{theorem}[MN-scheme as lifting construction]\label{th:mnrec}
Let $K$ and $t$ be positive integers such that $0<t<K$. 
Let 
\begin{align*}
    M_{K,0}([K]) =& J_{1,K}([K]), \qquad M_{K,K} = {\x}_{1,K} \\
    P_0 =& J_{\binom{K-1}{t},1}(\mathcal{S}), \qquad
    P_1 = M_{K-1,t-1}(\mathcal{S}),\\
    P_* =& M_{K-1,t}(\mathcal{S}_*),\qquad
    P_\# = \x_{\binom{K-1}{t-1},1},
\end{align*}
where $\mathcal{S}\cap\mathcal{S}_*=\emptyset$. Then, 
\begin{equation}
    M_{K,t}(\mathcal{S}\cup\mathcal{S}_*) = \mathcal{L}_{\{P_0,P_1\},\{P_*,P_\#\}}(\Tilde{I}_2)
\end{equation}
is a $(t+1)$-regular $(K,\binom{K}{t},\binom{K-1}{t-1},\binom{K}{t+1})$ PDA. 
\end{theorem}
\begin{IEEEproof}
     The set of PDAs $\mathcal{P}=\{P_0,P_1\}$ is  generalized Blackburn-compatible w.r.t. $\mathcal{P}_*=\{P_*,P_\#\}$.
     Also, we have $Z_1+Z_*=\binom{K-2}{t-2}+\binom{K-2}{t-1}=\binom{K-1}{t-1}$ and $Z_0+Z_\#=\binom{K-1}{t}$.
     Hence the lifting is valid due to Theorem~\ref{th:nonuni}.
     The number of integers in the lifted PDA is $|\mathcal{S}|=|\mathcal{S}_0|+|\mathcal{S}_*|+|\mathcal{S}_\#|=\binom{K-1}{t}+\binom{K-1}{t+1}+0=\binom{K}{t+1}$.
\end{IEEEproof}
\begin{example}
    We can use Theorem~\ref{th:mnrec} \nix{recursively} to construct $M_{4,2}$ as follows:
    \begin{align*}
    M_{2,1}(0) =& \mathcal{L}_{\{M_{1,0},J_{1,1}\},\{\x_{1,1},M_{1,1}\}}(\Tilde{I}_2) \\
        =& \begin{pmatrix}
            \x & M_{1,0}(0) \\
            J_{1,1}(0) & M_{1,1}
        \end{pmatrix}
        = \left(\begin{array}{c:c}
            \x & 0 \\ \hdashline
            0 & \x
        \end{array}\right) \\
        M_{3,1}([3]) 
        = &\begin{pmatrix}
            \x & M_{2,0}([2]) \\
            J_{2,1}([2]) & M_{2,1}(2)
        \end{pmatrix}
        = \left(\begin{array}{c:cc}
            \x & 0 & 1 \\ \hdashline
            0 & \x & 2 \\
            1 & 2 & \x
        \end{array}\right) \\
        M_{3,2}(0) 
        = &\begin{pmatrix}
            \x & M_{2,1}(0) \\
            J_{1,1}(0) & M_{2,2}
        \end{pmatrix}
        = \left(\begin{array}{c:cc}
            \x & \x & 0 \\ 
            \x & 0 & \x \\ \hdashline
            0 & \x & \x
        \end{array}\right) \\
        M_{4,2}([4]) &= \mathcal{L}_{\{J_{3,1},M_{3,1}\},\{M_{3,2},\x_{3,1}\}}(\Tilde{I}_2) \\
        = &\begin{pmatrix}
            \x & M_{3,1}([3]) \\
            J_{1,1}([3]) & M_{3,2}(3)
        \end{pmatrix}
        = \left(\begin{array}{c:ccc}
            \x & \x & 0 & 1 \\ 
            \x & 0 & \x & 2 \\ 
            \x & 1 & 2 & \x \\ \hdashline
            0 & \x & \x & 3 \\ 
            1 & \x & 3 & \x \\
            2 & 3 & \x & \x
        \end{array}\right)
    \end{align*}
\end{example}
A more general construction that subsumes the class of PDAs constructed using \cite{maddah2014fundamental} was proposed in \cite[Construction I]{shangguan2018centralized}. This construction also can be obtained using lifting. The construction needs the following result. 
\begin{lemma}\label{lm:shan}
    For integers $n,a,b$ such that $0<a,b<n$ and $n\ge a+b-1$, let $U_{n,a,b}$ denote the PDA constructed using \cite{shangguan2018centralized}. Then, $U_{n,a,b-1}$ and $U_{n,a-1,b}$ are right Blackburn-compatible w.r.t. $U_{n,a,b}$.
\end{lemma}
\begin{IEEEproof}
    Integers in both $U_{n,a,b-1}$ and $U_{n,a-1,b}$ are denoted by sets in $\binom{[n]}{a+b-1}$. Let an integer $s$, denoted by $\sigma\in\binom{[n]}{a+b-1}$, occur in the row denoted by $\rho\in\binom{[n]}{a}$ in $U_{n,a,b-1}$. Let $s$ also occur in the column $\gamma\in\binom{[n]}{b}$ in $U_{n,a-1,b}$. Then,
    \begin{align}
        \rho\cup\gamma\subseteq&\ \sigma,\\
        |\rho\cup\gamma|\le&\ a+b-1.\label{eq:rhogamma}
    \end{align} 
    For the entry $(\rho,\gamma)$ in $U_{n,a,b}$ to be an integer, we need $\rho$ and $\gamma$ to be disjoint and $|\rho\cup\gamma|=a+b$. Because of \eqref{eq:rhogamma}, this not true and the entry $(\rho,\gamma)$ is a $\x$.
\end{IEEEproof}

Construction~I from \cite{shangguan2018centralized} can be redefined as a recursive lifting construction as follows.
\begin{corollary}[Construction by Shangguan \textit{et al.} \cite{shangguan2018centralized}]\label{co:shan}
    Let $n,a,b$ be integers such that $0<a,b<n$ and $n\ge a+b-1$. Then
\begin{align*}
    U_{n,a,b}=\begin{cases}
        J_{\binom{n}{a},\binom{n}{b}}([\max (\binom{n}{a},\binom{n}{b})]) & \text{if } \min (a,b) = 0,\\
        \x_{\binom{n}{a},\binom{n}{b}} & \text{if } a+b=n+1,\\
        \mathcal{L}_{\{P_0,P_1\},\{P_*,P_\#\}}(\Tilde{I}_2) & \text{otherwise}
    \end{cases}
\end{align*}
is a $\binom{a+b}{a}$-regular $(\binom{n}{b},\binom{n}{a},\binom{n}{a}-\binom{n-b}{a},\binom{n}{a+b})$ PDA where the PDAs $P_0,P_1,P_*,P_\#$ are defined as follows:
\begin{align*}
    P_0 =& U_{n-1,a,b-1}(\mathcal{S}),\qquad
    P_1 = U_{n-1,a-1,b}(\mathcal{S}), \\
    P_* =& U_{n-1,a,b}(\mathcal{S}_*),\qquad
    P_\# = \x_{\binom{n-1}{a-1},\binom{n-1}{b-1}}.
\end{align*}
\end{corollary}
We skip the proof of Corollary~\ref{co:shan} as it is very similar to that of Theorem~\ref{th:mnrec}. An example is given below.

\begin{example}
Let $U_{n,a,b}$ be the PDA obtained from \cite{shangguan2018centralized} for integers $n,a,b$.
Let PDAs $P_0,P_1,P_*,P_\#$ be defined as follows:
\begin{align*}
    P_0 = U_{4,2,1}([4]) =& \begin{pmatrix}
\x & \x & 0 & 1 \\
\x & 0 & \x & 2 \\
\x & 1 & 2 & \x \\
0 & \x & \x & 3 \\
1 & \x & 3 & \x \\
2 & 3 & \x & \x \\
    \end{pmatrix}, \\
    P_1 = U_{4,1,2}([4]) =& \begin{pmatrix}
\x & \x & \x & 0 & 1 & 2 \\
\x & 0 & 1 & \x & \x & 3 \\
0 & \x & 2 & \x & 3 & \x \\
1 & 2 & \x & 3 & \x & \x \\
    \end{pmatrix},\\
    P_* = U_{4,2,2}(0) =& \begin{pmatrix}
\x & \x & \x & \x & \x & 0 \\
\x & \x & \x & \x & 0 & \x \\
\x & \x & \x & 0 & \x & \x \\
\x & \x & 0 & \x & \x & \x \\
\x & 0 & \x & \x & \x & \x \\
0 & \x & \x & \x & \x & \x
    \end{pmatrix},\\
    P_\# = \x_{4,4} =& \begin{pmatrix}
\x & \x & \x & \x \\
\x & \x & \x & \x \\
\x & \x & \x & \x \\
\x & \x & \x & \x \\
    \end{pmatrix}.
\end{align*}
The PDA $U_{5,2,2}$ can be obtained by lifting $\Tilde{I}_2$ with the above PDAs using Theorem~\ref{th:nonuni}.
\begin{align*}
    U_{n,a,b}&= \mathcal{L}_{\{P_0,P_1\},\{P_*,P_\#\}}(\Tilde{I}_2)\\
    &=\begin{pmatrix}
\x & \x & \x & \x & \x & \x & \x & 0 & 1 & 2 \\
\x & \x & \x & \x & \x & 0 & 1 & \x & \x & 3 \\
\x & \x & \x & \x & 0 & \x & 2 & \x & 3 & \x \\
\x & \x & \x & \x & 1 & 2 & \x & 3 & \x & \x \\
\x & \x & 0 & 1 & \x & \x & \x & \x & \x & 4 \\
\x & 0 & \x & 2 & \x & \x & \x & \x & 4 & \x \\
\x & 1 & 2 & \x & \x & \x & \x & 4 & \x & \x \\
0 & \x & \x & 3 & \x & \x & 4 & \x & \x & \x \\
1 & \x & 3 & \x & \x & 4 & \x & \x & \x & \x \\
2 & 3 & \x & \x & 4 & \x & \x & \x & \x & \x
    \end{pmatrix}.
\end{align*}
\end{example}

The following are a few more Blackburn-compatibility relations between MN PDAs.
\begin{lemma}\label{lm:il1}
The PDAs $\Tilde{M}_{K,t}$ and $M_{K,K-t-2}$ are right Blackburn-compatible w.r.t. $M_{K,K-t}$.
\end{lemma}
\begin{IEEEproof}
Both $\Tilde{M}_{K,t}$ and $M_{K,K-t}$ have $\binom{K}{t}$ rows. Rows of $M_{K,K-t}$ and $\Tilde{M}_{K,t}$ are labelled in lexicographic and reverse lexicographic order, respectively. So, the label for the $i$-th row in $M_{K,K-t}$ is the complement of the label for the $i$-th row in $\Tilde{M}_{K,t}$.
Columns of $M_{K,K-t-2}$ and $M_{K,K-t}$ are labelled with elements in $[K]$.
Also, both $\Tilde{M}_{K,t}$ and $M_{K,K-t-2}$ share the same set of integers $\mathcal{S}$ such that $|\mathcal{S}|=\binom{K}{t+1}$. Since the integers in $\Tilde{M}_{K,t}$ are labelled in reverse lexicographic order, label for an integer in $M_{K,K-t-2}$ is the complement of the label for the same integer in $\Tilde{M}_{K,t}$.

Let $s\in\binom{[K]}{t}$ be an integer in $\Tilde{M}_{K,t}$ and $M_{K,K-t-2}$. Let $s$ be labelled by $\tau$ in $\Tilde{M}_{K,t}$ and by $\sigma$ in $M_{K,K-t-2}$. We have,
\begin{align*}
    |\tau|=t,\qquad |\sigma|=K-t,\qquad \tau = [K]\setminus \sigma.
\end{align*}
Let $\Pi_s$ be the set of labels for rows in $\Tilde{M}_{K,t}$ containing integer $s$.
For $\pi\in\Pi_s$, we have $\pi\subset\tau$.
Let $\Pi_{*,s}$ be the set of labels for corresponding rows in $M_{K,K-t}$. We have $\Pi_{*,s}=\{[K] \setminus \pi: \pi\in\Pi_s\}$.
That is,
\begin{equation}
    \sigma\subset\pi_*,\qquad \forall \pi_*\in\Pi_{*,s}.
    \label{eq:il11}
\end{equation}
Let $\Gamma_s$ be the labels for columns in $M_{K,K-t-2}$ containing integer $s$.
We have,
\begin{equation}
    \gamma\subset\sigma,\qquad \forall\gamma\in\Gamma_s.
    \label{eq:il12}
\end{equation}
Combining \eqref{eq:il11} and \eqref{eq:il12}, we get,
$$
\gamma\subset\pi_*,\qquad \forall\gamma\in\Gamma_s, \forall \pi_*\in\Pi_{*,s}.
$$
Therefore, the entries in $M_{K,K-t}$ in the rows denoted by $\pi_*\in\Pi_{*,s}$ and columns denoted by $\gamma\in\Gamma_s$ are $\x$.
This proves the Blackburn-compatibility.
\end{IEEEproof}

Replacing $t$ with $K-t-2$ in Lemma~\ref{lm:il1}, we have the following corollary.
\begin{corollary}\label{co:il1}
The PDAs $\Tilde{M}_{K,K-t-2}$ and $M_{K,t}$ are right Blackburn-compatible w.r.t. $M_{K,t+2}$.
\end{corollary}
Using the above results, we can obtain the construction from \cite[Th.~4]{yan2017placement} for $M=\frac{1}{2}$.
Rather than lifting a base PDA, here we arrange PDA blocks next to each other such that Blackburn-compatibility is satisfied for blocks that share integers.

\begin{theorem}[\cite{yan2017placement}]\label{th:il1}
For integer $g>0$ and disjoint sets of integers  $\mathcal{S}_i, i\in [\lceil\frac{g+1}{2}\rceil]$ such that $|\mathcal{S}_i|=\binom{g}{2i}$, the array $P$ defined below is a $g$-regular $( 2g, 2^{(g-1)}, 2^{(g-2)}, 2^{(g-1)})$ PDA.
\begin{equation}
\resizebox{0.5\columnwidth}{!}{
    $P=\begin{pmatrix}
    \Tilde{M}_{g,1}(\mathcal{S}_1) & M_{g,g-1}(\mathcal{S}_0) \\
    \Tilde{M}_{g,3}(\mathcal{S}_2) & M_{g,g-3}(\mathcal{S}_1) \\
    \Tilde{M}_{g,5}(\mathcal{S}_3) & M_{g,g-5}(\mathcal{S}_2) \\
    \vdots & \vdots
    \end{pmatrix}.$}
\end{equation}
\end{theorem}
\begin{IEEEproof}
Since both $\Tilde{M}_{g,t}(\mathcal{S}_t)$ and $M_{g,g-t}(\mathcal{S}_0)$ have $\binom{g}{t}$ rows the array is well defined. Let $Z_0$ and $Z_1$ be the number of $\x$s per column in the two block columns, respectively. So, we have
$$Z_0=\sum_{\substack{t\in[g]\\ t~odd}} \binom{g-1}{t-1}=2^{g-2}.$$ 
If $g$ is even, then  
$$Z_1=\sum_{\substack{t\in[g]\\ t~odd}} \binom{g-1}{t-1}=2^{g-2}.$$
If $g$ is odd, then  
$$Z_1=\sum_{\substack{t\in[g]\\ t~even}} \binom{g-1}{t-1}=2^{g-2}.$$
Hence \ref{cond:equalZ} is proved. 
Let $\mathcal{S}=\bigcup_{i\in [\lceil\frac{g+1}{2}\rceil]} \mathcal{S}_i$. Since each block is a PDA, each integer appears in $P$, and this satisfies \ref{cond:everysOnce}. So,
$$|\mathcal{S}|=\sum_{\substack{t\in[g]\\ t~odd}} \binom{g}{t+1} = \sum_{\substack{t\in[g]\\ t~even}} \binom{g}{t}=2^{g-1}.$$ %
The same set of integers appear in $\Tilde{M}_{g,t}(\mathcal{S}_{(t+1)/2})$ and $M_{g,g-t-2}(\mathcal{S}_{(t+1)/2})$, and the Blackburn property holds for those integers due to Lemma~\ref{lm:il1} and Corollary~\ref{co:il1}. Since each integer occurs $(t+1)+(g-t-1)=g$ times, coding gain is $g$. Hence, $P$ is a $g$-regular $( 2g, 2^{(g-1)}, 2^{(g-2)}, 2^{(g-1)})$ PDA as claimed.
\end{IEEEproof}
\begin{example}
For $g=5$, Theorem~\ref{th:il1} results in the following $5$-regular $(10,16,8,16)$ PDA.
\begin{align*}
\resizebox{\columnwidth}{!}{$
    \begin{pmatrix}
    \Tilde{M}_{5,1}([1:10]) & M_{5,4}(0) \\
    \Tilde{M}_{5,3}([11:15]) & M_{5,2}([1:10]) \\
    \Tilde{M}_{5,5} & M_{5,0}([11:15]) 
    \end{pmatrix} 
    =
\left(\begin{array}{ccccc:ccccc}
    7  & 4  & 2  & 1  & \x  & \x  & \x  & \x  & \x  & 0  \\
    8  & 5  & 3  & \x  & 1  & \x  & \x  & \x  & 0  & \x  \\
    9  & 6  & \x  & 3  & 2  & \x  & \x  & 0  & \x  & \x  \\
    10 & \x  & 6  & 5  & 4  & \x  & 0  & \x  & \x  & \x  \\
    \x  & 10 & 9  & 8  & 7  & 0  & \x  & \x  & \x  & \x  \\ \hdashline
    12 & 11 & \x  & \x  & \x  & \x  & \x  & 1  & 2  & 3  \\
    13 & \x  & 11 & \x  & \x  & \x  & 1  & \x  & 4  & 5  \\
    14 & \x  & \x  & 11 & \x  & \x  & 2  & 4  & \x  & 6  \\
    15 & \x  & \x  & \x  & 11 & \x  & 3  & 5  & 6  & \x  \\
    \x  & 13 & 12 & \x  & \x  & 1  & \x  & \x  & 7  & 8  \\
    \x  & 14 & \x  & 12 & \x  & 2  & \x  & 7  & \x  & 9  \\
    \x  & 15 & \x  & \x  & 12 & 3  & \x  & 8  & 9  & \x  \\
    \x  & \x  & 14 & 13 & \x  & 4  & 7  & \x  & \x  & 10 \\
    \x  & \x  & 15 & \x  & 13 & 5  & 8  & \x  & 10 & \x  \\
    \x  & \x  & \x  & 15 & 14 & 6  & 9  & 10 & \x  & \x  \\ \hdashline
    \x  & \x  & \x  & \x  & \x  & 11 & 12 & 13 & 14 & 15
    \end{array}\right).$}
\end{align*}

\end{example}

\begin{table}[]
\centering
\caption{Comparison between Lemma~\ref{lm:liftbc} and existing constructions. 
Highlighted entries are shown in the plot in Fig.~\ref{fig:plotMR}.}\label{tab:comp}
\begin{adjustbox}{width=\columnwidth}
\begin{tabular}{cccccccc}
\toprule
Scheme & $g$ & $K$ & $f$ & $Z$ & $S$ & $\frac{M}{N}$ & $R$ \\ \midrule
Theorem~\ref{th:odd} & \multirow{2}{*}{11} & \multirow{2}{*}{22} & 22        & 19      & 6 & 0.8636 & 0.2727       \\
Lifting from \cite{aravind2022lifting}&&   & 22    & 20   & 4 & 0.9091 & 0.1818    \\\midrule[0.2pt]
Lemma~\ref{lm:liftbc} & \multirow{4}{*}{12} & \multirow{4}{*}{240} & 360        & 186      & 3480 & 0.5167 & 9.6667       \\
Lemma~\ref{lm:liftbc} & &        & 480        & 224      & 5120 & \cellcolor{teal2!50}0.4667 & \cellcolor{teal2!50}10.6667       \\
Lifting from \cite{aravind2022lifting}&&   & 960    & 588   & 7440 & \cellcolor{red2!70}0.6125 & \cellcolor{red2!70}7.75    \\
Cheng \textit{et al.} \cite{cheng2020some}&  &  & 64    & 60   & 80 & \cellcolor{yel2!70}0.9375 & \cellcolor{yel2!70}1    \\\midrule[0.2pt]
Huang \textit{et al.} \cite{huang2021new}& 10 & 240 & 64    & 48   & 384 & \cellcolor{cyan2!70}0.75 & \cellcolor{cyan2!70}6    \\\midrule[0.2pt]
Lemma~\ref{lm:liftbc} & \multirow{2}{*}{8} & \multirow{2}{*}{240} & 240        & 74      & 4980 & \cellcolor{teal2!50}0.3083 & \cellcolor{teal2!50}20.75       \\
Lifting from \cite{aravind2022lifting}&&  & 240    & 78   & 4860 & \cellcolor{red2!70}0.325 & \cellcolor{red2!70}20.25    \\\midrule[0.2pt]
Lemma~\ref{lm:liftbc} & \multirow{2}{*}{8} & \multirow{2}{*}{256} & 256        & 79      & 5664 & 0.3086 & 22.125       \\
Lifting from \cite{aravind2022lifting}&&   & 256    & 80   & 5632 & 0.3125 & 22    \\\midrule[0.2pt]
Lemma~\ref{lm:liftbc} & \multirow{2}{*}{16} & \multirow{2}{*}{256} & 256        & 147      & 1744 & 0.5742 & 6.8125       \\
Lifting from \cite{aravind2022lifting}&&   & 256    & 160   & 1536 & 0.625 & 6    \\ \bottomrule
\end{tabular}
\end{adjustbox}
\end{table}

\section{Results}\label{sec:results}
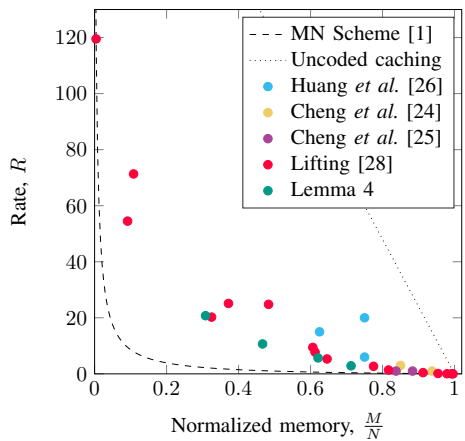
\begin{figure}[bth]
    \centering
    \definecolor{mycolor1}{rgb}{0.80000,0.94700,0.99100}%
\begin{tikzpicture}[scale=0.8]

\begin{axis}[%
clip mode=individual,
clip=true,
width=3in,
height=3in,
at={(0.758in,0.481in)},
xmin=0,
xmax=1.02,
ymin=0,
ymax=130,
axis background/.style={fill=white},
xlabel = {Normalized memory, $\frac{M}{N}$},
ylabel = {Rate, $R$},
legend style={legend cell align=left, align=left, draw=white!15!black}
]
    \addplot [domain=0:240, samples=241, black,dashed]({x/240},{(240-x)/(1+x)});
    \addplot[black,dotted] coordinates {(0,240) (1,0)};

\addplot[only marks, mark options={solid,draw=cyan2,fill=cyan2}] coordinates {(0.75, 6) (0.625, 15) (0.75, 20)};
\addplot[only marks, mark options={solid,draw=yel2,fill=yel2}] coordinates {(0.9375, 1) (0.85, 3)};
\addplot[only marks, mark options={solid,draw=purp2,fill=purp2}] coordinates {(0.88333, 1) (0.8375, 1)};

\addplot[only marks, mark options={solid,draw=red2,fill=red2
}] table[row sep=crcr]{%
x	y\\
0	240\\
0.004166667	119.5\\
0.1083333	71.33333\\
0.09166667	54.5\\
0.4833333	24.8\\
0.371875	25.125\\
0.325	20.25\\
0.60625	9.45\\
0.6125	7.75\\
0.6458333	5.3125\\
0.775	2.7\\
0.8166667	1.375\\
0.9125	0.4375\\
0.9541667	0.1375\\
0.9791667	0.04166667\\
0.9916667	0.0125\\
0.9958333	0.004166667\\
    };
    \addplot[only marks, mark options={solid,draw=teal2,fill=teal2}] coordinates {(0.3083, 20.75) 
    (0.4667, 10.6667) 
    (0.6208, 5.6875) 
    (0.7125, 2.875)
    };
   \addlegendentry{MN Scheme \cite{maddah2014fundamental}}
   \addlegendentry{Uncoded caching}
   \addlegendentry{Huang \textit{et al.} \cite{huang2021new}}
   \addlegendentry{Cheng \textit{et al.} \cite{cheng2020some}}
   \addlegendentry{Cheng \textit{et al.} \cite{cheng2021framework}}
   \addlegendentry{Lifting \cite{aravind2022lifting}}
   \addlegendentry{Lemma~\ref{lm:liftbc}}
\end{axis}
\end{tikzpicture}%
    \caption{Comparison of achievable $(M,R)$-pairs of low subpacketization schemes for $K=240$.}
    \label{fig:plotMR}
\end{figure}

\nix{For even ($K$), Theorem~\ref{th:il1} could obtain PDAs for $K$ users with $g=\frac{K}{2}$ with $\frac{M}{N}=\frac{1}{2}$. The MN construction would result in a PDA with coding gain $g=\frac{K}{2}$, when $t=\frac{K}{2}-1$, \textit{i.e.}, when $\frac{M}{N}=\frac{K-2}{2K}$. This value approaches $\frac{1}{2}$ as K increases. For large values of $K$, both MN construction and Theorem~\ref{th:il1} achieves coding gain $\frac{K}{2}$ at approximately the same memory ratio but our construction has much lower subpacketization compared to MN construction.}
Lifting constructions using Theorem~\ref{th:odd} and Lemma~\ref{lm:liftbc} are compared with the lifting schemes from \cite{aravind2022lifting} in Table~\ref{tab:comp}. 
In many scenarios, the proposed schemes achieve PDAs with same coding gain at a lower memory ($\frac{M}{N}$) compared to lifting schemes from \cite{aravind2022lifting}.
For $K=240$, $(M,R)$-pairs obtained using Lemma~6 is compared with those obtained in schemes with low subpacketization ($f<5K$) in the plot in Fig.~\ref{fig:plotMR}. 
A few $(M,R)$-pairs shown in the plot are highlighted in Table~\ref{tab:comp}.

\nix{
\begin{table}[]
\centering
\caption{Comparison between Theorem~\ref{th:il1} and MN construction.}\label{tab:comp}
\begin{tabular}{cccccc}
\toprule
Scheme & Gain, g & $K$ & $f$ & $Z$ & $S$ \\ \midrule
Theorem~\ref{th:il1} & \multirow{2}{*}{4} & 8       & 8        & 4      & 8       \\
MN&& 8  & 56    & 21   & 70    \\\midrule[0.2pt]
Theorem~\ref{th:il1} & \multirow{2}{*}{5} & 10      & 16       & 8      & 16      \\
MN&& 10 & 210   & 84   & 252   \\ \midrule[0.2pt]
Theorem~\ref{th:il1} & \multirow{2}{*}{6} & 12      & 32       & 16     & 32      \\
MN&& 12 & 792   & 330  & 924   \\ \midrule[0.2pt]
Theorem~\ref{th:il1} & \multirow{2}{*}{7} & 14      & 64       & 32     & 64      \\
MN&& 14 & 3003  & 1287 & 3432  \\ \midrule[0.2pt]
Theorem~\ref{th:il1} & \multirow{2}{*}{8} & 16      & 128      & 64     & 128     \\
MN&& 16 & 11440 & 5005 & 12870 \\ \bottomrule
\end{tabular}
\end{table}
}

\section{Conclusion}
\label{sec:conclusion}
We showed that lifting constructions is a versatile tool to design coded caching schemes and have the potential as a unifying framework. 
We proposed a few constructions for Blackburn-compatible PDAs. Lifting using these Blackburn-compatible PDAs improves the memory rate tradeoff for low subpacketization for a range of parameters.
Using a notion of generalized Blackburn-compatibility, we also brought some of the popular coded caching schemes into the ambit of lifting constructions. 
The existence of more non-uniform Blackburn-compatible PDAs could be explored in future work.

\section{Acknowledgment}
V. R. Aravind is supported by the 5G Testbed Project funded by the Department of Telecommunications, Government of India.

\end{document}